\magnification=1200

\centerline {\bf CONFORMALLY  INVARIANT $\sigma$ MODELS  ON $AdS$ SPACES , }
\centerline{\bf  CHERN-SIMONS $p$-BRANES, HADRONIC BAGS  AND  $W$ GEOMETRY } 
\bigskip
\centerline { Carlos Castro}
\centerline { Center for Theoretical Studies of Physical Systems}
\centerline { Clark Atlanta University, Atlanta, GA. 30314, USA}
\bigskip
\centerline { Revised October, 2000}
\bigskip

\centerline{\bf ABSTRACT}
\bigskip

Conformally invariant sigma models in $D=2n$ dimensions with target non-compact
$O(2n,1)$ groups are studied. It is shown that despite the non-compact
nature of the $O(2n,1)$ groups, the classical action and Hamiltonian are positive
definite. Instanton field configurations are found
to correspond geometrically to conformal ``stereographic''  mappings
of $R^{2n}$ into the Euclidean signature $AdS_{2n}$
spaces. Zaikov's relationship between Self Dual $p$-branes and Chern-Simons
$p'$-branes, provided $p=p'+1$ and the embedding $D=p+1$-dimensional 
manifold is Euclidean , is elaborated further. 
Branes actions can be obtained also from a Moyal deformation quantization of Generalized Yang Mills
Theories. Using this procedure, we show how four dimensional 
$SU(N)$ YM theories contain Chern-Simons membranes and hadronic bags in the large $N$ limit.  
Since Chern-Simons $p'$-branes have
an underlying infinite dimensional algebra containing $W_{1+\infty}$ ,
as shown by Zaikov, we discuss the importance that $W$ geometry should have in the final 
formulation  of $M$ theory.

\bigskip

\centerline{\bf I. INTRODUCTION    }
\bigskip 

Conformally invariant  $O(2n+1)~\sigma$ models in $2n$ dimensions were
of crucial importance in the construction of conformally invariant
Lagrangians, with vanishing world-volume cosmological constant,  for bosonic $p$-branes ( $p+1=2n$ ) [1,2]. 
In particular, Self Dual unit charge instanton solutions were found which correspond to conformal ( stereographic) maps from $R^{2n}\rightarrow S^{2n}$. These models [1] were the higher-dimensional extension of the $O(5)~\sigma$ 
models of Feisager and Leinaas in $4D$ [3].   When $p+1=odd$, the authors [2] also built Lagrangians for bosonic extendons ( branes) , however, conformal invariance was lost. These sort of Lagrangians allowed the author to construct a $polynomial$ action for the spinning membrane after a Weyl covariantization process was performed [4]. Another Weyl invariant spinning membrane action was constructd by [5] but it was 
$non-polynomial$ complicating the quantization program. 

Conformal Field Theories have  risen to more prominence recently mainly due to Maldacena's conjecture on the 
AdS/CFT duality between M/string theory  on    
$AdS_d\times S^{D-d}$ backgrounds and CFT's living on the projective boundary of the $AdS_d$ spaces [6]. Relevant $\sigma$ models with target spaces on certain supergroups have been used to decribe 
$CFT$ on $AdS$ backgrounds with Ramond-Ramond (RR) 
Flux [7]. In particular, various exact $2D~CFT$ on $AdS_{2n+1}$ backgrounds have been constructed 
recently that could be used in building superstring theories propagating on $AdS_{2n+1}$ backgrounds [8]. These $\sigma$ models were based on the standard $SL(2,R)$ WZNW model; i.e 
$\sigma$ models on group manifolds with WZNW terms. 

Conventional $\sigma$-models are based on compact groups like $O(2n),
SU(n)...$. In general, compact simple groups are used mainly to
simplify the quantization program; i.e the 
quantization is not riddled with the standard problems of 
ghosts  due to the non-positive definite inner products; uniqueness
of  the WZ functional; solvability of the model; positivity of the
Hamiltonian [29,30]. The authors [29] studied $O(N,1),SU(N,1)$ non-compact sigma models
in two-dimensions and have shown that a dynamical mass generation with asymptotic
freedom is possible and that a sensible unitary quantization program
is  possible by recurring to  a selection
rule in the Hilbert space of states ( only postive definite
norms are allowed).

Setting aside these technical subtleties on the compact simple character of the group,  and motivated by the recent findings on Anti de Sitter spaces,  we will study conformally invariant 
( noncompact ) $O(2n-1,2)~\sigma$-models in $2n$ even dimensions; i.e maps from  
$R^{2n}$ to $ O(2n-1,2)$ and $O(2n,1)$. Instanton solutions are found
for the latter case, corresponding 
to conformal, ``stereographic `` maps from $R^{2n}$ to $ O(2n,1)$. 
The $SO(2n,1)$ group manifold, $modulo$ the action of its maximal compact subgroup $SO(2n)$, $SO(2n,1)/SO(2n)$,   
is topologically the Euclidean signature $AdS_{2n}$ whose natural isometry group is $SO(2n,1)$.  
To be more specific, we shall concentrate on $O(2n-1,2)$ and 
$ O(2n,1)~\sigma$-models keeping in mind that the relevant groups are the $SO(2n-1,2)$ and
$SO(2n,1)$.

The  Lorentz signature $AdS_{2n}$ space can be viewed 
as a hyperboloid embedded in a pseudo-Euclidean $2n+1$-dim manifold
with coordinates $y^A=y^0, y^1,...,y^{2n}$ and diagonal metric given
by 
$\eta_{AB}=diag~(-,+,+,..,+,-)$ with length squared $y^Ay_A$ preserved by the isometry group $SO(2n-1,2)$.  
The  $AdS_{2n}$ is  defined as the geometrical locus  :

$$y^Ay^B\eta_{AB} =-R^2 =-1. \eqno (1)$$
de Sitter spaces require a change in the sign in the r.h.s.

Signature subtleties are crucial in the construction of instanton solutions. The ordinary Hodge dual star operation is $signature$ dependent. For example, the double Hodge dual star operation acting on a rank $p$ differential form in a $4D$ space satisfies :

$$^{**}F=s(-1)^{p(4-p)}F.~~s=+1 ~for~(4,0); (2,2).~s=-1~for~(3,1); (1,3)\eqno (2)$$
where we have displayed the explicit signature dependence in the values of $s$. Hence, for a rank two form $F$, in Euclidean $4D$ space and with signature  $(2,2)$ ( Atiyah-Ward spaces) one can find solutions to the (anti) self dual YM equations : $ ^* F=\pm F$. There are no YM instantons in $4D$ Minkowski space since in the latter one has imaginary eigenvalues  : $ ^* F=\pm  i F$.

The $two~temporal$ variables are required in the embedding process of
$AdS_{2n}$ into the pseudo-Euclidean space  $R^{2n-1,2}$  We will
show, in fact, 
that it is the noncompact  $O(2n,1)~\sigma$-model, instead of the 
$O(2n-1,2)~\sigma$-model , that has instanton solutions obeying a double self duality condition similar to the one obeyed by the BPST instanton [9].
It was shown [1] that , the Euclidean ( compact) $O(2n+1)~\sigma$ model instanton solutions are directly related to instanton solutions of $O(2n)$ Generalized YM (GYM) theories in $R^{2n}$ [10] obeying the forementioned double self duality condition.  

In view of the conformally invariant $\sigma$ models/GYM connection in
higher dimensions, 
the next step is to study $p$-branes  ( with $p+1=2n$ ) 
propagating on $AdS_D$ backgrounds. In particular, the critical case
when the dimensionality of the target space is saturated , $D=p+1=2n$.
Dolan-Tchrakian [1,2] constructed the corresponding conformally
invariant  Skyrme-like actions, with vanishing world-volume
cosmological constant , based on these conformally invariant $\sigma$ models in $2n$-dimensions. Upon the algebraic elimination of the auxiliary world-volume metric, Dolan and Tchrakian have shown that one recovers the Dirac-Nambu-Goto action :

$$S=T\int d^{2n} \sigma~\sqrt { |det~G_{\alpha \beta}| }. 
~~~G_{\alpha,\beta} =\eta_{\mu\nu}\partial_\alpha X^\mu \partial_\beta X^\nu. \eqno (3)$$
where $G_{\alpha \beta} $ is the induced world-volume metric resulting
from the embedding of the $p+1=2n$ hypervolume into the $D$-dim
target spacetime. 
When the spacetime
dimension is saturated : $D=p+1=2n$ the square root of the
Dirac-Nambu-Goto action simplifies and one obtains the usual Jacobian
for the change of variables from $\sigma $ to $X$.  In such case the Nambu-Goto action is topological  :   there are no physical local transverse degrees of freedom . Such  topological  actions have been studied by  Zaikov [11]. For a  review of membranes and other $p-branes$ with extensive references see [12]. The action is then :

$$ S= T \int d^{2n} \sigma~ 
\partial_{\sigma^1} X^{\mu_1}\wedge ....... \wedge \partial_{\sigma^{2n} } X^{\mu_{2n}}. \eqno (4)$$
where $T$ is the $p$-brane ( extendon) tension. 
For a $p$-brane  whose world-volume has a natural boundary , an integration ( Gauss law) yields :

$$ S= T \int_{\partial V} d^{2n-1} \zeta~ 
X^{\mu_1}\wedge \partial_{\zeta^1} X^{\mu_2}\wedge ....... \wedge \partial_{\zeta^{2n-1} } X^{\mu_{2n}}. \eqno (5)$$
one then recovers the action for the Chern-Simons $p'$ brane whose $p'+1$ world-volume variables , 
$\zeta^a,~a=1,2,.....p'+1$, are  integrated over the 
$2n-1$-dim  boundary $\partial V$ of the $2n$-dim domain $V$
associated with the world volume of the open $p$-brane. The value of
$p'$ must be such that $p'+1=p=2n-1$  [11]. Zaikov concluded that
these topological Chern Simons $p'$-branes exit only in target
spacetimes of dimensionality $D=p'+2$. 

In particular, when the dimensionality of the target spacetime is
saturated , $D=p+1$, one can construct, in addition , self-dual
$p$-brane (extendon) solutions obeying the equations of motion and
constraints ( resulting from $p+1$ reparametrization invariance of the
world-volume) that are directly related to these topological
Chern-Simons $p'$-branes. This holds provided $p'+1=p$ and the embedding manifold is Euclidean [12]. Furthermore, 
when $D=p+1=2n$ one has conformal invariance a well [1,2]. It is in
this fashion how the relationship between the self dual $p$-branes and
Chern Simons $p'=p-1$ branes emerges . This is roughly the analogy with Witten's discovery of the one-to-one relationship between $3D$ nonabelian Chern-Simons theories and $2D$ rational CFT. [13]. 

The topological Chern-Simons action (5) and the action (4) , obtained
by using Gauss law , are among the main  ones considered in this
work.

In section {\bf II } we will show that the Euclidean signature $AdS_{2n}$  signature is an 
instanton solution of 
the $O(2n,1)~\sigma$-models. In {\bf 3.1} we discuss the relation between ( open ) Self Dual 
$p$-branes and Chern-Simons $p'$-branes for $ p = p'+1$. 
In {\bf 3.2} we show that ( spacetime filling ) $p$-branes can be obtained from 
a Moyal deformation quantization ( of the Lie algebraic structure ) of Generalized Yang  Mills theories [1] ( GYM).
Adding a topological term makes these theories $nontrivial$ due to boundary dynamics.  
In particular, we briefly review [35] how Chern-Simons membranes and Hadronic Bags emerge 
from the large $N$ limit of quenched, reduced $4D$ $SU(N)$ YM  (QCD ) [36] . 
In {\bf 3.3} {\cal W} geometry is discussed in connection to these
theories. Finally we present our conclusions.

\bigskip
\centerline{\bf II Euclidean  $AdS_{2n} $ as $O(2n,1)~\sigma$-model
Instantons }
\bigskip
As a prototype we will imagine a Chern-Simons membrane living on  the Euclideanized $AdS_4$ 
background , which in turn, will be shown to be the instanton field configuration of 
the conformally invariant $O(4,1)~\sigma$ model in $R^4$ obeying the
double self-duality condition to be described below by eq-(8b).

Since the conformally invariant Euclidean $O(5) \sigma$-model in
$R^4$ has a correspondence with the conformally invariant $O(4)$ YM in
$R^4$ [1] it is natural to ask whether the analytical continuation  from $O(4,1)\rightarrow O(5)$ will allow  to establish the following relationships in four-dimensions among : 
topological Chern-Simons membrane living in ( Euclidean) $AdS_4$; instanton
configurations of the conformally invariant $O(4,1)~\sigma$-models in
$R^4$ and $O(4)$ Yang-Mills instantons ( obeying a double self duality
condition described below by eq-(8a)) in $R^4$.

The importance to establish this web of relationships is because one may generalize this web to higher dimensions with the provision that $D=2n=4k$. 
For example, in $D=12$ we may have  the connections among the
Chern-Simons $p=10$-brane moving on  the Euclideanized $AdS_{12}$;
instantons of the conformally invariant
$O(12,1)~\sigma$-models in  $R^{12}$ and $O(12)$ Generalized
Yang-Mills (GYM) theories in $R^{12}$  . When the group is compact like $O(2n)$ , for example, the latter GYM are defined by Lagrangians [1,10] in $R^D$ where $D=2n=4k$ :

$$L=tr~(F^{\alpha_1 \alpha_2....\alpha_{2k}}_{\mu_1,\mu_2....\mu_{2k}}
~\Sigma_{\alpha_1,\alpha_2.....\alpha_{2k}})^2 = (F_{2}\wedge
F_{2}....\wedge F_{2})^2. ~~~2n=4k=D\eqno (6)$$
with all indices antisymmetrized. $\Sigma_{\alpha_1 \alpha_2...}$ is the totally antisymmetrized product ( on the internal indices 
$\alpha$) of the product of $k$ factors of the $2^{2k-1}\times 2^{2k-1}$ matrices $\Sigma_{\alpha_1 \alpha_2}$
corresponding to the chiral representation of $SO(4k)$.

The exhibiting relationship between the instanton field configurations of the $O(5)~\sigma$ model and the $O(4)$ YM system in $R^4$ were given by [1] . The order parameter field of the $O(5)~\sigma$ model is the $O(5)$ vector $n^i (x)$ obeying the constraint :$n^i (x) n_i (x) =1$ . The definition of the gauge fields and field strengths for the YM  system in terms of the $n^i(x)$ are :

$$ A^{ij}_\mu =n^i (x) \partial _\mu n^j (x) - n^j (x) \partial _\mu n^i (x) .                                                   \eqno (7a)$$

$$F^{ij}_{\mu \nu} (x) =\partial _\mu A^{ij}_\nu - A^{ik}_\mu
A^{lj}_\nu \eta_{kl} - \mu \leftrightarrow \nu = \partial_{[\mu } n^{
  [ i} \partial_{ \nu ]} n^{j ]}.                                      \eqno (7b)$$
due to the constraint $n^in_i=1$ the former field stength coincides
with the definition given below by eq-(8c). When $n^i n_i = -1$ one should
have $+A^{ik}_\mu A^{lj}_\nu$ in the definition (7b) instead; i.e. a
change in sign in the coupling constant. These signs subtleties are
essential otherwise the field strength will be zero. 

An additional constraint allows to reduce $O(5)\rightarrow O(4)$ : 

$$(\delta ^i_j \partial_\mu + A^{ij}_\mu ) n^i (x) =0                                                  \eqno (7c)$$

After using eqs-(7a,7b,7c) the authors [1] have shown that $O(4)$ YM  instanton solutions obey a double self duality condition satisfied by the BPST instanton [9] :   :

$$ \epsilon_{\alpha_1 \alpha_2 \beta_1 \beta_2} 
F^{\beta_1 \beta_2 }_{\mu_1 \mu_2} = \epsilon_{\mu_1 \mu_2 \nu_1\nu_2} F^{\nu_1 \nu_2}_{\alpha_1 \alpha_2}\eqno (8a)$$
contrasted with the double self-duality condition exhibited by the $O(5)~\sigma$ model  in $R^4$ [1] :

$$ \epsilon_{i_1i_2.....i_5} n^{i_5} (x)  
F^{i_3 i_4 }_{\mu_1 \mu_2} = \epsilon_{\mu_1 \mu_2 \nu_1\nu_2} F^{\nu_1 \nu_2}_{i_1 i_2}                                      \eqno (8b)$$
that are related  to the $\sigma$ model Lagrangian in $D=4$
$$S=\int d^4x~{1\over 2 (2!)} (F^{ij}_{\mu \nu})^2.~ ~ ~
F^{ij}_{\mu \nu} = \partial_{[\mu}  n^i (x)\partial _{\nu]} n^j (x) 
-\partial_{[\mu}  n^j (x) \partial _{\nu]} n^i (x).\eqno (8c)$$

Similar type of actions are generalized for higher rank objecs :
$$S=\int d^{2n}x~{1\over 2 (n!)} (F^{i_1...i_n}_{\mu_1....\mu_n})^2.~ ~ ~
F^{i_1....i_n}_{\mu_1...\mu_n} = \partial_{[\mu_1}  n^{[i_1} (x).....\partial
_{\mu_n]} n^{i_n]} (x) \eqno (8d)$$

The unit charge instanton solutions of eq-(8b) that minimize the action (8c) and correspond to the $O(4)$ YM instanton solutions of eq-(8a) were found to be precisely the ones corresponding to the stereographic projections ( conformal mappings) from $R^4 \rightarrow S^4$ [1] :
$$n^a (x) = {2 x^a \over 1+x^2}; a=1,2,3,4. ~~~n^5 (x) ={x^2-1 \over
  x^2 +1}.
~~~\sum_{a=1}^{a=4} (n^a)^2 + (n^5)^2 =1. \eqno (9)$$

Things differ now for the (noncompact )$O(2n,1)~\sigma$-models due to the subtle signature changes. Now we must see whether or not the $O(2n,1)$ vector $n^i(x)$ with $i=1,2,......2n+1$ satisfy the double self duality condition 
 $$\epsilon_{i_1......i_n i_{n+1}..........i_{2n}.i_{2n+1}} n^{i_{2n+1}} F^{i_{n+1} i_{n+2}.....i_{2n}}_{\mu_{n+1}.........\mu_{2n}} = \epsilon_{\mu_1 \mu_2.....\mu_{n+1}........\mu_{2n} }  F^{\mu_1 \mu_2.......\mu_{n} }_{i_1 i_2......i_n}      \eqno (10)$$

The ``stereographic'' maps from $R^4$ to the Euclidean signature 
$AdS_4$ defined by :

$$n^a (x) ={2 x^\mu \over 1-x^2};~ a=1,2,3,4.~\mu =0,1,2,3. ~~~n^5 (x) ={1+x^2 \over 1-x^2}. \eqno (11a)$$
$$x^2 \equiv (x^0)^2 +(x^1)^2 + (x^2)^2+(x^3)^2.
     \eqno (11b)$$
obey the condition : 

$$n^i n_i   \equiv (n^1)^2 +(n^2)^2 + (n^3)^2+(n^4)^2 -(n^5)^2 =-1. \eqno (11c)$$
which is just the $O(4,1)$ invariant norm
of the vector $n^i(x)=(n^1(x),.....n^5(x))$  
and satisfy the double-self duality condition conditions (8b) as we
shall show in this section . It is important to emphasize that the double self duality conditions are not derived from 
the actions (8c, 8d). However, solutions to the double self duality conditions automatically 
obey the equations of motion, due to the analog of the Bianchi Identities, like it occurs in ordinary YM theories. 
But the converse is not true.

Another important issue we shall  address now is the issue of singularities. The solutions 
given by Eqs-(11) $obey$ the double self duality conditions and 
are naturally singular at the points $x^2 = 1$. These points have a correspondence  with the 
projective/conformal boundaries at infinity as we will show at the end of this section. 
Nevertheless, eqs-(8a, 8b) are still satisfied and no singularities occur in the definition of the norm. 
These singularities are naturally due to the noncompact nature of $AdS$ spaces.  

The most famous example is the $AdS$ metric : the $AdS$ conformally flat metric is singular 
at the points $x^2 = 1 $ as well. The scalar curvature is well defined and equals the 
negative cosmological constant $ R = \Lambda $ as it should. However, the 
Einstein-Hilbert action associated with the $AdS$ metric $blows ~up $ due to the 
divergent  contributions of the determinant of the metric; i.e the measure of integration diverges 
at $x^2 = 1$ :

$$ g_{\mu\nu} =  { 1 \over ( 1 - x^2 )^2 } \eta_{\mu\nu} \Rightarrow   
\int d^{2n}x \sqrt{ g } R =  \int d^{2n}x \sqrt{g} \Lambda = 
\Lambda \int d^{2n}x  { 1 \over ( 1 - x^2 )^{2n} }. \eqno ( 12) $$
Using the standard power counting arguments one can show that  the integral $diverges$ due to the pole at $x^2 = 1$. 
The fact that the action $diverges$ is not a reason to $abandon$ the $AdS$ metric solutions to the equations 
$ R = \Lambda$. By the same token, despite the singularities of the instanton field configurations due 
to the non-compact nature of $O(2n,1)$, one should not disregard them as being irrelevant nor pathological.

A similar reasoning can be applied to the Schwarzschild solution  that has a true physical singularity 
at the origin $r = 0$. And also to the Poincare metric associated with the projective disk model in the 
complex plane. The latter is singular at the boundary of the disc which can be holomorphically be mapped 
 to the real axis of the complex plane.

These singular field configurations have `` zero measure `` contribution in the Euclidean  path integral
since they are exponentially supressed $ e^{ -S_{E}} \rightarrow 0$.   
At the classical level, one can regularize their singular contributions in the classical action.  
This is achieved by adding compensating surface terms to the bulk actions
(8c, 8d) and in this fashion one can cancel out the divergent contributions due to the points $x^2 = 1 $.
These boundary terms are the analogs of the $\theta$ topological terms in YM theories.  
Notice that the actions in eqs-(8c,8d) have the YM form : $ F\wedge ^ * F$. The topological $\theta$ terms 
have the form : $\theta F \wedge F $ where by in our case by the $*$ operation one means the double star operation 
with respect to the base manifold indices and target space indices and is given up to factorial numerical factors by :  

$$   ^*[~F^{\nu_1 \nu_2.....\nu_n}_{i_1 i_2.....i_n }~] 
\equiv  \epsilon^{\mu_1 \mu_2.....\mu_n \nu_1\nu_2......\nu_n} \epsilon_{i_1i_2.....i_{2n+1}} n^{i_{2n+1} } (x)  
F^{ i_{n+1} i_{n+2} ....i_{2n} }_{\mu_1 \mu_2.....\mu_{n}}. \eqno (13) $$

The singular field configurations (11) are nothing but the solutions to the double self duality equations , therefore 
by adding the analog of a theta term ( with the $suitable$ sign )  of the form : $\theta ~F\wedge F $,  
to the YM-type $bulk$ action $ F \wedge ^* F $ , given by eqs-(8c,8d) and  
evaluated at the singular field configurations (11),  one will be able to cancel out 
those singular contributions to the bulk action due to the behaviour at $x^2 = 1 $ ; i.e to the projective 
boundaries at infinity.  

Another alternative is to simply introduce a natural `` cutoff ``  in defining/restricting 
the $domain$ of integration by imposing $x^2 = \lambda^2 < 1$  The analog of this procedure 
is the selection of the fundamental modular domain of the string one loop ( multiloop ) path integrals. 
The Siegel upper half complex plane is noncompact with infinite area. 
Dividing by the action of the discrete modular group one ends up by having one single copy of the 
fundamental domain which is still noncompact but has a $finite$ area. 
An analogous example of this ( although not identical ) is the action for a point
particle. The  action naturally diverges if one takes the interval of proper time from $0$ to $\infty$. 
The cut-off simply states to define the action in the compact domain where the proper time is bounded $0, \tau$.

Before we finalize the discussion on the issue of singularities it is very important to $emphasize$ that 
the solutions to the double self duality conditions are defined $ modulo$ conformal  transformations 
of the base space $R^{2n}$. One could re-absorb the singular behavour of the instanton solutions at $x^2=1$ 
inside these  conformal scaling factors. For example, the $AdS$ metric is singular at $x^2 = 1$ 
because the conformal factor
is precisely singular at $x^2 = 1$. The $AdS$ metric is conformally flat and therefore belongs in the same 
conformal class as the flat metric. Under this procedure the conformally re-scaled field configurations 
will be well behaved , however the action (8c, 8d) will still diverge due to the singular behaviour of the 
determinant of the new re-scaled metric. This is similar  to what occurs with the Einstein-Hilbert action 
associated with the $AdS$ metric : the scalar curvature $ R = \Lambda $ is finite 
but the action diverges due to the singularity in the determinant of the metric at $x^2 =1$.

After this discussion on adding a compensating boundary term to the actions (8c, 8d) , ...etc....
if one wishes to regularize the divergent contributions due to the natural singularities at $x^2 =1$,  which 
correspond to the projective boundaries at infinity, we will proceed to study the signature
dependence in the definition of the duality operation.  
For the particlar case  when $2n=4$, the double duality star operation
is defined   :

$$^{*} F\equiv \epsilon^{\mu_1 \mu_2 \nu_1 \nu_2}\epsilon_{i_1 i_2 i_3 i_4
  i_5} n^{i_5} F^{i_3 i_4}_{\mu_1 \mu_2}. \eqno (14)$$
The $^{**} $ acting on $F$ is :

$$^{**} F=(-1)(-1)^{s_b}(-1)^{s_g}(-1)^f F. \eqno (15)$$
where (i) the first factor of $-1$ stems from the $n^i n_i =-1$
condition. (ii) The second factor stems from the signature of
the base space $R^4 $. (iii) The third factor stems from the group
signature which is $(-1)^1=-1$ for $O(4,1)$. (iv) The last one,
$(-1)^f$,  are
additional factors resulting from the permutation of the indices in
the $\epsilon_{\mu_1....\mu_{2n}}$ and $\epsilon_{i_1....i_{2n+1}}$
tensors . These are similar to  the 
$(-1)^{r(D-r)}$ factor appearing in the double Hodge operation. Where
$r$ represents the rank of $F$. 
In this case they yield the factor $1$. 

From (15) one immediately concludes that the signature of the base
manifold must be : $(+,+,+,+)$ so that  
 
$$^{**} F=(-1)(-1)^{s_b}(-1)^{s_g}F=(-1) (+1) (-1)(1) F=F\Rightarrow
(^{*})^2 =1 \Rightarrow ^{*} =\pm 1 . \eqno (16a)$$
and one has a well defined double (anti) self duality condition.  
$$^{*} F=\pm F. \eqno (16b)$$

For higher rank fields, $F^{i_1.....i_n}_{\mu_1.....\mu_n}$ it follows
from (15) that  we must have in addition that $n=even$. Therefore we conclude that
the base manifold must be of Euclidean signature type and have for dimension: $R^{4k}$
and the target group background $O(4k,1)$,  whose topology is
that of the Euclideanized $AdS_{4k}$ . Thus, the maps are
signature-preserving and the dimensionality of the base manifold must
be a multiple of four,  $D=4k$ . 

It still remains to prove that the $O(4k,1)~\sigma$-field $n^A (x)$ given by eqs-(11) solve the double self
duality condition (10); that they saturate the minimum of the action to
ensure topological stability, if
indeed such minima exit, which is not necessary the case. Also the instanton solutions must yield a
finite valued action; i.e the fields must fall off sufficiently fast
at infinity so the action does not blow up;....

To satisfy these requirements it is essential to prove firstly that
the classical action and classical Hamiltonian are positive
definite. A close inspection reveals that one should have an Euclidean
signature base manifold which forces one to choose an Euclidean
signature $AdS_{2n}$ space if the condition (16a) is to be
satisfied. Therefore, instead of $O(2n-1,2)$ one must have $O(2n,1)$.  
To illustrate the fact that the classical action is positive definite
and without loss of generality we study the $R^{4,0} \rightarrow
AdS_{4,0}$ case. ( the Euclideanization of $AdS_4$).  We must show
that :

$$(F^{ij}_{\mu\nu})^2 >0.~i,j =1,2,3,4,5.~\mu, \nu =0,1,2,3.  \eqno (17)$$
In $ D=4$ there are six terms of the form : 

$$g^{00}g^{11}[ (F^{ab}_{01})^2 - (F^{5a}_{01})^2 ] +
g^{00}g^{22}[ (F^{ab}_{02})^2 - (F^{5a}_{02})^2 ] +.....+$$
$$g^{11}g^{22}[ (F^{ab}_{12})^2 - (F^{5a}_{12})^2 ] +.......\eqno (18)$$
Due to the Euclidean nature of the base manifold the $g^{00}g^{11}>0$
so we
must show then that the terms inside the brackets in eq-(18) are all positive
definite. The $a,b$ indices run over : $1,2,3,4$ and the last one is
the $i=5$ component $n^5$ associated with the non-compact
$O(4,1)$-valued $n^i$ field. 
Setting $n^i = (n^a, n^5 )\equiv (\vec \pi, \sigma)$ the condition
$n^i n_ i = -1$ yields :
$$\sigma^2 = \vec \pi. \vec \pi +1 =\pi^2 +1. \eqno (19)$$
which implies for every single component of $x^\mu; \mu =0,1,2,3$ ( we
are not summing over the $x^\mu$) :

$$(\partial_\mu n^5)^2 < {\pi^2 \over \sigma^2 }(\partial_\mu \vec
\pi).(\partial_\mu \vec \pi);~\mu =0,1,2,3. \eqno (20)$$
Then :
$$(F^{5a}_{01})^2 = {1 \over \sigma ^2}[ (\partial_0  \pi^a )(\vec
\pi . \partial_1 \vec \pi) - (\partial_1  \pi^a )(\vec
\pi . \partial_0 \vec \pi) ]^2 <{\pi^2 \over \sigma^2 }
(\partial _0 \pi^a \partial_1  \pi^b - \partial _1 \pi^a \partial_0
\pi^b) ^2. \eqno (21)$$
where we used  the relation $(\vec A. \vec B)^2 \le A^2 B^2$. The first
term of (18) yields :
$$(F^{ ab}_{01})^2 - (F^{5a}_{01})^2> (1-{\pi^2 \over \sigma^2 })  (\partial _0 \pi^a \partial_1  \pi^b - \partial _1 \pi^a \partial_0
\pi^b) ^2 >0. \eqno (22)$$
because $(1-{\pi^2 \over \sigma^2 })>0$ due to the relation (19). 
The same argument applies for each single component of $(F^{ij}_{\mu
\nu})^2$ and hence the classical action and Hamiltonian are positive
definite. This is an essential ingredient for the instanton solution to saturate
the minimum of the classical action. From ordinary Yang-Mills in Euclidean space
and for $compact$ gauge groups like $SU(2)$ we have the familiar relation
: 

$$\int d^4 x~ (F- ^* F)^2 =2\int d^4 x~ F^2 - 2 \int d^4 x ~F ^* F
\ge 0 \Rightarrow \int d^4 x ~F^2 \ge \int d^4 x ~F ^* F  \eqno (23)$$
in the self dual case, the last equality holds. The last integral
for the $SU(2)$ group ( topologically $S^3$) is
the winding number $S^3 \rightarrow S^3$, a topological invariant; hence the self dual
instanton obeying $F = ^* F$ saturates the lower bound of the positive
definite action and ensures the topological stability. To finalize one
must have that the field strengths fall off sufficiently fast at
infinity to ensure that the action doesn't blow up : 
$ r^n F^{i_1...i_n}_{\mu_1....\mu_n} \rightarrow 0$ as $r\rightarrow \infty$.

In the non-compact $AdS$ case studied here one naturally encounters 
singularities at $x^2 = 1$ for the instanton configurations corresponding to the 
projective boundaries at infinity. 
The classical action for the $O(4,1)~\sigma$-model in
$R^4$ was positive definite and since the instanton solutions saturate the lower
bound of the classical action this implies that the action diverges. The source of its 
divergence is naturally due to the $noncompact$ nature of the $O(2n, 1)$ group. To regularize 
the action one can add compensating boundary terms and/or restrict the domain of integration 
to a compact region $inside$  the $2n$-dimensional disc $ x^2 = 1$.

Whatever the choices to regularize the action may be this does not change the crux of this work :  
The instanton solutions found  in eqs-(11) are $genuine$ solutions of 
the double self duality conditions (8b,13) which are an integrable  subset of the 
equations of motion for the conformally invariant $\sigma$-models 
on $AdS_{2n}$ spaces given by the action of eqs-(8c,8d).  
The instanton solutions are defined $modulo$ a conformal transformation of the base manifold 
$R^{2n}$.

It remains to prove that the solutions (11) obey the double self
duality condition (10). This former
property is a straightforward extension of the
results in [1,3] for the compact case. Given
the stereographic  projections defined by eq-(11) : $R^{2n}
\rightarrow H^{2n}$ one can
find a set of independent $2n+1$ frame vectors, $E_p$,  representing the pullback of the
orthogonal frame vectors of the $2n+1$-dim pseudo-Euclidean manifold onto the
Euclidean signature $AdS_{2n}$ space  :

$$E_1=\partial_{x^1} n^i; ~E_2=\partial_{x^2} n^i;..... 
E_{2n}=\partial_{x^{2n}}
n^i.~E_{2n+1}=n^i. ~ i=1,2,.....2n+1\eqno (24)$$
obeying  the orthogonality condition : 

$$g_{\mu\nu}\equiv E^i_\mu E^i_\nu =\partial_\mu n^i \partial_\nu n^i
={4\over (1-x^2)^2} \eta_{\mu\nu}. \eqno (25)$$
which precisely yields the Euclideanized $AdS_{2n}$ metric $g_{\mu\nu}$ definded as
the pullback of the embedding metric, satisfying $R_{\mu\nu}\sim
g_{\mu\nu}\Lambda$ and $R\sim \Lambda$. The latter curvature relations
are the hallmark of Anti de Sitter space ( negative scalar curvature). 
The positive definite measure spanned by the collection of $2n+1$ frame vectors 
defines the natural volume element of the Euclideanized $AdS_{2n}$ :

$$\sqrt {det ~h} = det~(E_1,E_2,......,E_{2n}, E_{2n+1} ). \eqno (26)$$

It was shown by [1,3] (for the compact case the orthogonality
condition is naturally modified to yield the metric of the $S^{2n}$) that 
the orthogonality conditions eqs-(25) are
the $necessary$ and $sufficient$ conditions to show that the
particular field configurations $n^i(x)$ given by
(11), after extending the four dimensional result to any $D=2n$,
satisfy the double self duality conditions eq-(10). 
The orthogonality relation implies that the angles among any two vectors is
preserved and that all vectors $after$ the conformal mapping has been taken have equal length 
relative to each other ( this does not mean that they have the same length as before ! ). 

This is a signal of a conformal mapping. The ``stereographic'' maps are
a particular class of conformal maps. The action of the conformal
group in $R^{2n}$ will furnish the remaining conformal maps [1,3]. 
Hence, we see here once again the interrelation between self-duality
and conformal invariance. In two dimensions self-duality implies
conformal invariance ( holomorphicity). In higher dimensions matters
are more restricted, they are not equivalent.

Therefore, setting aside the subtleties in the quantization program
due to the 
noncompact nature of $O(2n,1)$,  we have shown that these
$O(2n,1)~\sigma$-model instanton field configurations obtained by
means of the ``stereographic'' ( conformal) maps of $R^{2n}
\rightarrow H^{2n}$, correspond precisely to the coordinates,
$y^1,y^2,....y^4,y^5$ of the  $5$-dim pseudo-Euclidean manifold onto
which the Euclidean signature $AdS_4$ space is embedded. 
To illustrate this lets focus on Euclidean signature $AdS_4$ defined as  the geometrical locus :

$$y^2 \equiv (y^1)^2 +(y^2)^2 + (y^3)^2+(y^4)^2 -(y^5)^2 =
-\rho^2=-1. \eqno (27)$$
where we set the $AdS$ scale $\rho=1$. 
The $AdS_4$ coordinates : $z^0,z^1,z^2,z^3$ are related to the $y^A$
using the ``stereographic'' projection ( see the lectures by Petersen
[6]) from the ``south'' pole of the Euclidean signature $AdS_4$ to the equator :

$$y^a  (z) =\rho{2 z^\mu \over 1-z^2};~\mu=0,1,2,3. ~a=1,2,3,4. 
~~~y^5 (z) =\rho {1+z^2 \over 1-z^2}. \eqno (28)$$
it is straighforward to verify that $y^2=-\rho^2=-1$. 
The Euclidean signature $AdS$ corresponds to two hyperbolic branches. A north and south branch respectively
intersecting the vertical axis at $ y^5 = \pm \rho $. The mappings are performed by projecting 
the rays stemming from the `` south pole `` $S$,   the vertex of the southern    hyperbolic    
branch located at $ y^5 = - \rho$,  
intersecting the equator $R^{2n}$ ( passing through the origin dividing the north and south
branches ) at point $P'$ . One extends the rays $ SP'$ through the equator upwardly 
until they intersect the upper hyperbolic branch at point $P$.  
It is in this fashion how one constructs the `` stereographic `` mappings (28).

The Euclidean signature $AdS_4$ metric is given :

$$g_{\mu \nu} = {4 \over (1-z^2)^2} \eta_{\mu \nu}¸~~~ z^2 =(z^0)^2
+....+(z^3)^2. \eqno (29)$$ 

As mentioned above, the metric blows up at $x^2 =1$, points that correspond to the projective boundaries 
at infinity.  
As we can verify by inspection, there is an $exact$  match with the instanton field configurations $n^i
(x)$ (11). The metric (29) also matches (25) 
after the correspondence : $n^a (x)\leftrightarrow y^\mu (z)$  and
$n^5 (x) \leftrightarrow y^5 (z)$ is made. 

Therefore, the main conclusion of this section is : 
The instanton field configurations of the conformally 
invariant $O(2n, 1)~\sigma$-models, obeying the 
double self duality conditions (8b,13 ) correspond geometrically to the conformal `` stereographic'' maps 
of $R^{2n}$ into the Euclidean signature $AdS_{2n}$ spaces.

These conformally invariant $O(2n,1)$ sigma model actions in $R^{2n}$
are just the sigma model generalizations of the MacDowell-Mansouri action for
ordinary gravity based on gauging the conformal group $SO(4,2)$ in
$D=4$ Minkowski space and the Anti de Sitter group $SO(3,2)$. 
The ordinary Lorentz spin connection and the tetrad in four dimensions
are just pieces of the $SO(3,2)$ gauge connection. Upon setting the
torsion to zero one recovers the $4D$ Einstein-Hilbert action with a
cosmological constant and the Gauss-Bonnet topological term. 
The conformally invariant sigma models actions [1] are the ones required in building
conformally invariant bosonic $p=2n$-brane actions in flat or curved target bcakgrounds. It would be
interesting to see what sort of actions can be derived following the
analog of a
MacDowell-Mansouri procedure to obtain Einstein gravity from a gauge theory .

\bigskip
\centerline{\bf III. } 
\bigskip 

In this section we shall discuss in detail the connection between 
Self-Dual $p$-branes, Chern-Simons $p'$-Branes ( for $ p = p'+1$) and ${\cal W} $ Geometry. 
We will also  ouline the straightforward steps to derive 
(spacetime-filling) $p$-branes from Moyal Deformation Quantization of 
Generalized Yang Mills Theories [1, 34 ] . 
In particular, we shall show how Chern-Simons membranes and Hadronic bags 
$emerge$  from Quenched Large $N$ QCD  . We leave all the details for the references [34, 35]. 
\bigskip

\centerline{\bf 3.1 Self Dual $p$-branes }
\bigskip

As pointed in the introduction, Zaikov noticed that self-dual $p$-branes, when $p+1=D=2n$,
are related to Chern-Simons $p'$-branes ( $p'+1=p$) provided the
embedding manifold  is  Euclidean  [11]. The
Euclidean $Ads_{2n}$ space can be seen, not only as a
$O(2n,1)~\sigma$-model instanton, but also  as a self-dual $p$-brane whose
world volume is the $AdS_{2n}$ space ($p+1=2n=D$ dimensional).  
To see this we must first write down Dolan-Tchrakian action for a bosonic
$p$-brane with $p+1=2n$ embedded in a flat/curved target space of
dimensionality $D=p+1=2n$. Dolan-Tchrakian action is valid for flat or
curved target spacetime metrics [2]. 

$$S = T \int d^{2n}\sigma~
[E^{p_1}\wedge E^{p_2}\wedge...\wedge E^{p_n}]\wedge ^{*}
[(E^{q_1}\wedge E^{q_2}\wedge....\wedge E^{q_n})]\eta_{p_1 q_1}\eta_{p_2q_2}..... \eqno (30)$$
the action is YM like $F\wedge ^{*} F$ and was derived based on  the
conformally invariant $O(2n+1)~\sigma$-model actions. The star
operation in (30) is the standard Hodge duality one defined w.r.t the $p+1$-dim
world-volume metric of the $p$-brane : $h_{ab}$. The world-volume
one-forms ,
$E^p;~p=1,2,....2n$ are obtained as the pullbacks of the $D=2n=p+1$ one
forms, $e^A~;A=1,2,....2n$ 
associated with the $D=2n$ dimensional flat/curved target space onto
which we embed the
world-volume of the $p$-brane. A self-dual $p$-brane ( self dual w.r.t the Hodge star
operation) obeys :

$${*}
[(E^{q_1}\wedge E^{q_2}\wedge....\wedge E^{q_n})]=[E^{p_1}\wedge
E^{p_2}\wedge...\wedge E^{p_n}]. \eqno (31)$$
a self-dual $p$-brane will automatically satisfy the equations of
motion. This is the $p$-brane generalization of the relation
$d^{*}F=^{*}j=0$ in the absence of sources in the YM equations. When
$^{*}F=F$ the latter equations become the Bianchi identities $dF=0$. 
The action (30) for a self-dual $p$-brane becomes then the integral
over the ordinary Jacobian ( $p+1=2n$ volume form) associated with the change
of variables $\sigma \rightarrow X$ and hence is topological. If the
world volume has a natural boundary an integration by parts ( Gauss
law) will yield the Chern-Simons $p'=p-1$-brane. In this way we can
see once more how the
self-dual $p$-brane is related to the Chern-Simons $p'=p-1$-brane ( at
least on shell).  

This is the first step. One needs to show now that the self-duality
condition (31) is directly related to the self-duality condition
formulated by Zaikov when $p+1=2n$ and, furthermore, that it is also
related to the double-self duality conditions (10) associated with the
conformally invariant $O(2n,1)~\sigma$-models in $R^{2n}$. The self-duality
condition formulated by Zaikov when the embedding manifold is
Euclidean and when $p+1=4$, for example,  is [11] : 

$${\cal F}^{JK}_{ab}= \partial_{\sigma^a} X^J \partial_{\sigma^b} X^K-\partial_{\sigma^b} X^J \partial_{\sigma^a} X^K
={1\over 4}\epsilon _{abcd}\epsilon^{JKLM} {\cal F} ^{LM}_{cd}. \eqno
(32)$$
where $X^I (\sigma^a)$ are the embedding coordinates of the $p$-brane
into the target space, 
$X^I;~I=1,2,...2n$. Eq-(32) is the self-duality condition for case
when $p=3~; p+1=4$ obtained from Dirac-Nambu-Goto actions. The
Dolan-Tchrakian action (30) is equivalent at the classical level to
the Dirac-Nambu-Goto actions upon elimination of the auxiliary
world-volume metric, $h_{ab}$. Whether or not this equivalence occurs Quantum
Mechanically is another issue. Hence, classically, Zaikov's
self-duality conditions (32) are equivalent to the self-duality
conditions (31) of the self-dual extendon furnished by Dolan-Tchrakian
actions (30).

Now we must proceed with the connection between the latter
self-duality for extendons ($p$-branes) and the double self
duality conditions associated with the $O(2n,1)~\sigma$-models.  The self-duality condition of Zaikov has exactly
the same form as the double self-duality condition for the $O(4)$ YM
instanton (8a), obtained from the compact $O(2n+1)~\sigma$ model,  after using the
constraint  (7c), $D_\mu (A)n^i=0~; i=1,2,...2n,2n+1$ ( for the particular
case $n=2$ ), and the relations (7a,7b) which allow to reduce
$O(5)$ to $O(4)$ and show that the $O(4)$ YM field is a $composite$
field expressed in terms of the $n^i$.   We
need now to establish the relations/correspondences  between  the $p$-brane coordinates
$X^I (\sigma^a)$ of (32)  and the $O(4)$ YM $composite$ fields of eqs-(7a,7b)
appearing in eq-(8a). 
The ensuing relation is :

$$A^{\alpha \beta}_\mu \Sigma_{\alpha \beta} \leftrightarrow X^J (\sigma^a). ~~~F^{\alpha \beta}_{\mu \nu} \Sigma_{\alpha \beta}
\leftrightarrow 
{\cal F}^{JK}_{ab}= \partial_{\sigma^a} X^J \partial_{\sigma^b}
X^K-\partial_{\sigma^b} X^J \partial_{\sigma^a} X^K. \eqno (33)$$
where $\Sigma_{\alpha \beta}$ are the $2\times 2$ matrices ( Pauli)
corresponding to the chiral representation of $SO(4)\sim SU(2)\times
SU(2)$. The $\mu,\nu..$ indices in the l.h.s of (33) run over the
four dimensional base manifold $R^4$ where the $O(4)$ YM fields
live. The $I,J,K...$ in the r.h.s of (33) run over the
four-dimensional embedding target space . The $\sigma^a$ indices run over the
four-dimensional world volume of the $p=3$-brane. 
  
The last equation is reminiscent of the chiral model approaches to
Self Dual Gravity based on Self Dual Yang Mills [31] theories. A Moyal
deformation quantization of the Nahm equations associated with a $SU(2)$ YM [32] theory yields the $classical$~$N\rightarrow
\infty$ limit of the $SU(N)$ YM  Nahm equations  $directly$,  without ever using the 
$\infty \times \infty$ matrices of the large $N$ matrix models. By simply taking the
classical $\hbar =0$ limit of the
Moyal brackets, the ordinary Poisson bracket algebra associated with
area-preserving diffs algebra $SU(\infty)$   is  automatically recovered.   
This supports furthermore the fact that the underlying geometry may be
noncommutative [18,32] and that $p$-branes are essentially gauge theories of
area/volume...preserving diffs [14,20,21].     

The way this is attained is the following. A Moyal quantization takes
the operator ${\hat A}_\mu (x^\mu)$ into $A_\mu(x^\mu;q,p)$ and
commutators into Moyal brackets. A dimensional reduction to one
temporal dimension brings $A_\mu (t,q,p)$, which precisely corresponds
to the membrane coordinates $X_\mu (t,\sigma^1,\sigma^2)$ after
identifying $\sigma^a$ with $q,p$. The $\hbar =0$ limit turns the Moyal
bracket into a Poisson one. It is in this
fashion how the large $N~ SU(N)$ matrix model bears a relation to the 
physics of membranes. The Moyal quantization explains this in a
straightforward fashion without having to use $\infty \times \infty$
matrices !

Therefore, the self-duality conditions of the Dolan-Tchrakian
$p$-brane actions  in Euclidean embedding manifolds ( for the case
$p+1=4$)  (31) have a one-to-one correspondence with
Zaikov's self-duality conditions (32) ( for Euclidean embedding
manifolds) and with the double
self-duality conditions of the $O(4)$ YM instanton (8a), if, and only
if, the $gauge~fields/target~space~coordinates $ correspondence (33)
is used.

Finally, if the Euclidean signature $AdS_{2n}$ space is to be seen as a
self-dual $p$-brane; i.e. as the Euclideanized world-volume of an open  self-dual $p$-brane
such $p+1=2n$, in addition to an $O(2n,1)~\sigma$-model instanton in $R^{2n}$, we
must analytically continue $O(2n,1)$ to $O(2n+1)$ and , afterwards,
follow the same procedure as above, relating the $O(5)~\sigma$-model
to $O(4)$ YM and its Generalized Yang-Mills extensions in $D=2n=4k$
dimensions [10]. For example, one could first reduce $O(4,1)$ to
$O(3,1)$ using equations like (7a,7b,7c) and
perform aftewards the analytical continuation $O(3,1)$ to $O(4)$ . It
is important to have a compact group like $O(4)$ otherwise we will run
into the same problems of non-positive definite actions, Hamiltonians... 

To sum up, the arguments ranging from  eqs-(31-33), are the
$necessary$ ones
required that would allow
us to view the Euclidean $AdS_{2n}$ space as the Euclideanized world volume
of an open  self-dual $p$-brane ( $p+1=2n$) in addition to being an
$O(2n,1)~\sigma$-model instanton . In section {\bf 3.2} we will present further 
evidence which suggests that 
there is a formal equivalence between the two definitions of
duality ; i.e after using the gauge
fields/coordinates correspondence (33) one may be able to show that the  
 Euclidean $AdS_{2n}$ space is  the Euclideanized world volume
of an open  self-dual $p$-brane ( $p+1=2n$) in addition to being an
$O(2n,1)~\sigma$-model instanton.

After having presented the discussion of the self-dual $p$-branes /Chern-Simons
$p'=p-1$-branes relationship in target spacetimes of dimension 
$D=p+1=p'+2=2n=4k$, we will  choose a particular higher dimensional example. 
The topological Chern-Simons action for the $p=10$-brane is defined over 
the $d=11$-boundary domain of $R^{12}$ . Boundary to-boundary maps are a special class, 
among the infinite family of embeddings of the Chern-Simons $p=10$-brane into $AdS_{12}$,   
which  conformally map the $d=11$-dim boundary domain into the projective boundary of the $AdS_{12}$ 
space : which is   topologically  $S^1\times S^{10}$. 

The Euclidean signature $AdS_{12}$ has, instead,  for projective
boundary the $S^{11}$. The signature matching condition (16a) imposes the
signature of the base manifold to be Euclidean . If one views  $R^{12}$ 
as the interior of a twelve-dimensional  ball, $B_{12}$ of very large
($\infty$) 
radius, the $d=11$ boundary domain is effectively $S^{11}$
. Therefore, for the very special case associated with boundary
to-boundary  maps,   the Chern-Simons 
$p=10$-brane can be characterized by the winding number of the mappings of
$S^{11} \rightarrow S^{11}$ . In this special case, the
Chern-Simons $p=10$-brane lives effectively on the  projective boundary  of
Euclidean $AdS_{12}$. We emphasize that this is just a particular
case,  since in general,  the topological Chern-Simons $p=10$-brane
lives on the $whole~bulk$  of $AdS_{12}$. Since the Chern-Simons
$p=10$-brane theory is topological,
it has no local bulk degrees of freedom, only global, that are naturally
confined to the boundary. In this sense the theory is holographic
[27,28].

Furthermore, since $AdS_{12}$ can be seen as a hyperboloid embedded in a $d=13$-dim pseudo-Euclidean manifold we have  $d=11,12,13$ for characteristic dimensions; hence this construction may be  relevant to understand  the intricate relations of $M,F,S$ Theory  [25,26] and to shed some light into the geometrical, holographic and topological underpinnings behind the AdS/CFT duality conjecture. 

A Chern-Simons $p=11$-brane living on a Euclidean signature
$AdS_{13}$ 
embedded in a $d=14$-dim pseudo Euclidean space has for world
volume a $12$-dim manifold. In this case the relevant dimensions would
be $11,12,13,14$. The main difference is that the $odd$ dimensional
$AdS_{13}$ will $not$ corespond to an instanton configuration
associated with an $O(13,1)~\sigma$-model because in odd dimensions
conformally invariant sigma models of the Dolan-Tchrakian type cannot
be constructed. 
Higher dimensional topological actions have been proposed by Chapline
to describe unique theories of gravity and matter [24]. 
\bigskip
\bigskip
\bigskip
\bigskip

\centerline{\bf 3.2  Branes from Moyal Deformation Quantization of GYM Theories }
\smallskip
\centerline{\bf Chern Simons Membranes and Hadronic Bags from Quenched large N QCD } 
\bigskip

$SU(N)$ reduced, quenched gauge theories have been shown to be related to string theories in the large 
$N$ limit. We will review very briefly the steps that show how a $4D$ YM reduced, quenched, theory supplemented 
by a topological $\theta$ term can be related, through a Weyl Wigner Groenowold Moyal (WWGM) quantization procedure, 
to an open $3-brane$ model [35] . The bulk is the interior of a hadronic bag and the boundary is the  
Chern-Simons world volume of a membrane. The boundary dynamics {\bf is not } trivial despite the fact that there are no 
transverse bulk dynamics associated with the interior of the bag.
For further details we refer to [35] . 

The reduced and quenched YM action in $D = 4 $ is  [36] : 

$$ S = -{1 \over 4} ({ 2 \pi \over a } )^4 { N \over g^2_{YM} } ~ Tr~ F_{\mu\nu} F^{\mu\nu}. ~~~ 
F_{\mu\nu} = [ i D_\mu, i D_\nu ]. \eqno (34) $$
Notice that the reduced quenched action is define at a `` point `` $x_o$.  
For simplicity we are omitting the matrix $SU(N) $ indices. 

The $\theta$ term is :

 $$ S = -{\theta N g^2_{YM}  \over 16 \pi^2 } ({ 2 \pi \over a } )^4 
\epsilon^{\mu\nu\rho\sigma}Tr~ F_{\mu\nu} F_{\rho \sigma}. \eqno (35)$$

The WWGM quantization establishes a one-to-one correspondence bewteen a linear operator 
$D_\mu $ acting on the Hilbert space ${\cal H} $ of square integrable functions on $R^D$ and a smooth 
function ${\cal A}_\mu ( x, y )$  which is the inverse Fourier transform of   ${\cal A}_\mu ( q, p )$. The latter is 
obtained by evaluating the Trace of the operator $D_\mu $ by means of summing over the diagonal elements 
with respect to an orthonormal basis in the Hilbert space ${\cal H} $. Under the WWGM correspondence 
the matrix product turns into the Moyal $*$ product and the commutator turns into the Moyal bracket :

$$ { 1 \over \hbar } [ A_\mu, A_\nu ] \rightarrow \{ {\cal A}_\mu, {\cal A}_\nu \}.  \eqno( 36) $$

The WWGM deformation of the actions is :

$$ {\cal S}  = -{1 \over 16 } ({ 2 \pi \over a } )^4 ({ N \over 2 \pi })^2  { 1 \over g^2_{YM} } 
\int d^{2D}\sigma ~ {\cal F_{\mu\nu}}*{\cal F} ^{\mu\nu}. 
 \eqno (37) $$

The corresponding WWGM deformation of the $\theta$ term is :

 $$ {\cal S}_{\theta}  = -{\theta N g^2_{YM}  \over 64  \pi^2 } 
({ 2 \pi \over a } )^4   ({ N \over 2 \pi })^2 
\int d^{2D}\sigma~ \epsilon^{\mu\nu\rho\sigma}{\cal F} _{\mu\nu}*  {\cal F}_{\rho \sigma}. \eqno (38)$$
where the Moyal mappings of the initial field strenghts are explicit $ \sigma$ dependent ( phase space ) 
quantities : $ {\cal F} _{\mu\nu} ( \sigma )$ where $\sigma$ runs over the $2D$-dimensional phase space spanned by 
$ q^i, p^i;~ i = 1,2,3 ...D$.

By making the gauge fields/coordinates correspondence :

$$ {\cal A}_\mu ( \sigma ) \rightarrow ( { 2\pi \over N} )^{1/4} X_\mu ( \sigma ).~~~ 
{\cal F}_{\mu\nu}  ( \sigma ) \rightarrow ( { 2\pi \over N} )^{1/2 }  \{ X_\mu ( \sigma ), X_\nu (\sigma)  \} \eqno ( 39a ) $$
performing the correspondence bewteen the Trace and phase space integration :  
$$ { (2\pi)^4  \over N^3 } Tr  \Rightarrow \int d^{2D} \sigma . \eqno ( 39b)$$ 
and by taking the deformation parameter $''\hbar''$ of the WWM quantization to be  $2\pi / N = \hbar = 0$ 
then the `` classical limit `` is nothing but the equivalent of the large $N$ limit. Hence, the 
 Moyal brackets turn into 
Poisson brackets, the factors of $N$ decouple and the  
first action (37) is just equivalent to the Dolan-Tchrakian action evaluated in the conformal gauge.
The second action (38) is the 
Chern-Simons Zaikov action for a membrane whose world volume lives on $3$-dim the boundary of the hadronic bag.
Upon identifying the inverse lattice spacing of the large $N$ quenched, reduced  {\bf QCD} with the QCD scale 
of $(2\pi /a) = \Lambda_{QCD} = 200 ~Mev$ we obtained the value of the dynamically generated bag pressure 
$\mu_o  $ given by :

$$ \mu_o^4 = { 1 \over 4 \pi } ( 200 ~ Mev)^4 { 1\over ( g^2_{YM}/4\pi)  }. \eqno(40a)  $$
taking for the conventionally assumed value of $ g^2_{YM}/4\pi  \sim 0.18 $ we get an actual value for the 
hadronic bag constant $ \mu_o $ close to the value of $\Lambda_{QCD} = 200~ Mev$ which falls within the range of the 
phenomenologically accepted values of $\mu_o \sim 110~ Mev$,  $if$  we take into account the uncertainty on the 
value of the  $\Lambda_{QCD} $ which lies in the range of :  $ 120~Mev < \Lambda_{QCD} < 350~Mev$.  
One should notice also the interesting fact. If one sets the QCD scale $ a$ to be :

$$ a^4 = N L^4_{ Planck} ~ in ~ \mu_o^4 \sim { 1 \over a^4  g^2_{YM}}  \Rightarrow 
\mu_o^4 \sim { 1 \over  L^4_{ Planck}(N g^2_{YM}) }. \eqno (40b)$$
one makes contact with Maldacena's celebrated result which relates the size of the Anti de Sitter space 
throat $ a^4$ with the 
't Hooft coupling $ N g^2_{YM}$ and the Planck scale  $L^4_{ Planck}$ ( Gravity ) . 
The large $N$ limit, or strong coupling limit, 
is implemented by the double scaling relation $ L^4_{ Planck} N = a^4 $ finite as $ N \rightarrow \infty$ and 
$ L^4_{ Planck} \rightarrow 0$. 
Hence, relations (40a, 40b)  very naturally reflect Maldacena's result which support his $AdS/CFT$ duality conjecture 
obtained from a completely different approach that did not require compactifications. 
True, we need to work further to show why $ a^4 = N L_{Planck}^4 $ , howgravity emerges. 
Since we obtained brane actions from YM ( GYM) theories 
via the Moyal deformation quantization and branes $do$ contain gravity 
it is not surprising to arrive at similar results.

To extend these results to higher $p$-branes than $ p = 3 $ one needs to perform the Moyal deformation of 
Generalized Yang Mills Theories (GYM) as shown by [1, 34, 35 ] .
The GYM  are based on $SO( 4k ) $ in $R^{D}$ where $ D = 4k $ : 

$$L = { 1\over g^2 } ~ tr ~ ( F^{\alpha_1 \alpha_2........\alpha_{2k} }_{\mu_1\mu_2....\mu_{2k}} 
\Sigma_{ \alpha_1 \alpha_2........\alpha_{2k}} )^2 . \eqno (41)$$
where $g$ is a dimensionless coupling constant and : 

$$\Sigma_{ \alpha_1 \alpha_2........\alpha_{2k}} = \Sigma_{\alpha_1 \alpha_2}  \Sigma_{\alpha_3 \alpha_4}...\eqno (42)$$
is an anti-symmetrized product of $ k$ factors of the $2^{2k -1}\times 2^{2k-1} $ matrices 
$\Sigma_{\alpha\beta}$ corresponding to the chiral representation of $SO(4k)$ . For $ k = 1$ one has the usual 
$SO(4)\sim SO(3)\otimes SO(3)$ whose double cover is $SU(2)\otimes SU(2)$. For further details we refer to [1,34]. 

The analogous procedure of quenching and reduction is attained by looking only at the zero mode sector of the theory. 
One can write down the action for those field strength configurations that are everywhere constant. 
A local gauge transformation 
will rotate them to other field strength configurations that are space-time dependent. 
Since the invariant action requires performing a a group trace   
to attain gauge invariance,  one may simply focus, from the start, on 
those field strength configurations that are everywhere 
constant. Performing a WWGM deformation of the Lie algebraic structure in each single one of the subfactors 
 $\Sigma_{\alpha\beta}$ appearing in the definition of the $SO(4k)$-valued field strenghts and integrating out the 
spacetime dependence one arrives [34] : 

$${\Omega_{4k} \over g^2 } \int d^{4k}\sigma ~ [~ \{ A_{\mu_1}, A_{\mu_2} \} \{ A_{\mu_3}, A_{\mu_4} \}
.......                  \{ A_{\mu_{2k -1} }, A_{\mu_{2k} } \}                ]^2. \eqno (43)$$

where $  \Omega_{4k} $ is the spacetime volume resulting from the spacetime integration of constant field configurations 
( zero modes ). Taking the $\hbar = 0$ classical limit and establishing the 
gauge fields/coordinates $ A_\mu \rightarrow (X_\mu / l_p) $   correspondence ( units require dividing by the 
Planck scale $l_p$ ) one arrives at the follwoing $p$-brane action and Tension :

$${\Omega_{4k} \over g^2 l_p^{8k}} \int d^{4k}\sigma ~ [~ \{ X_{\mu_1}, X_{\mu_2} \} \{ X_{\mu_3}, X_{\mu_4} \}
.......\{ X_{\mu_{2k -1} }, X_{\mu_{2k} } \}                ]^2.~~~ T_p =  {\Omega_{4k} \over g^2 l_p^{8k}}.  \eqno (44)$$
Notice that the fields in the reduced-quenched $SU(N) $ QCD action do not have the canonical dimensions. 
They are rescaled [35]. The action (44) is equivalent to the DT action [34] in the conformal gauge. 
As before, this bulk action does not have transverse degrees of freedom : 
one simply can choose the orthonormal gauge 
to see that this action reduces to a pure bulk volume term.
equivalent to the DT action after choosing the coformal gauge.      
However, the analog of a the topological $theta$ terms  are $not$ trivial. The have true boundary dynamics 
and these correspond naturally to the Chern-Simons $p'$-branes discussed earlier. Following the same steps as we did 
for the $theta$ terms in the reduced, quenched large $N$ QCD action one can verify that the Lagrangian density 
for the $\theta$ terms :

$$L_{\theta} \sim { \theta~ g^2 } ~  \epsilon^{\mu_1 ....\mu_n \nu_1....\nu_n} 
tr ~ ( F^{\alpha_1 \alpha_2........\alpha_{2k} }_{\mu_1\mu_2....\mu_{2k}} 
\Sigma_{ \alpha_1 \alpha_2........\alpha_{2k}} )   
( F^{\beta_1 \beta_2........\beta_{2k} }_{\nu_1\nu_2....\nu_{2k}} 
\Sigma_{ \beta_1 \beta_2........\beta_{2k}} )               . \eqno (45)$$

will yield the Zaikov Chern-Simons $p'$-brane actions after performing the WWGM deformation of the 
Lie algebraic structure. Upon  taking the $\hbar = 0$ limit; identifying the traces with 
integrals over the phase space variables which later are identified with  the world volume coordinates , 
and establishing the gauge field/coordinates correspondence one ends up with  :

$$ S_{CS} \sim  {\Omega_{4k} \theta g^2 \over  l_p^{8k}} 
\int d^{4k}\sigma ~ dX_{\mu_1} \wedge dX_{\mu_2} \wedge.........\wedge   dX_{\mu_{4k} }   . \eqno (46)$$
after an integration by parts that will yield the Zaikov Chern-Simons $p'$- brane action integrated 
over the $ 4k -1$-dim boundary so that the value of $p'$ must obey $p'+1 = 4k -1$.

Having shown that certain brane actions and the associated Chern-Simons branes, 
can be obtained from a Moyal deformation of the Lie algebraic bracket structure of the YM ( GYM) theories 
and that Chern-Simons membranes and Hadronic Bags 
emerge from a reduced, quenched large $N$ QCD in four dimensions, for example,  allows to confirm once more    
what have established in this work :

$\bullet$ Topological Chern-Simons $p$-Branes and Conformally Invariant $\sigma$ models on $AdS$ spaces are related 
and connected  with $SO(4k)$ Generalized YM
theories in $R^{4k}$ ( where $k = 2n $). 

$\bullet$ We have closed  the web
among three related theories : (i) Chern-Simons $p$-branes on $AdS_{2n}$
spaces with $p+2=2n$;  (ii) $O(2n,1)$ Conformally
invariant $\sigma$ models in $R^{2n}$ with target Euclideanized $AdS_{2n}$ backgrounds and (iii ) obtained  branes 
from a Moyal deformation of 
$SO(4k)$ GYM theories in $R^{2n} = R^{4k} $ . 

$\bullet $  Also , the relationship
between self duality and conformal invariance has been established : The Euclidean signature
$AdS_{2n}$ space was shown to correspond precisely to instanton field
configurations of the noncompact $O(2n,1)~\sigma$-models in $R^{2n}$,
obeying the double self duality condition (10), which was the analog
of the BPST instanton [9].   
\bigskip
\centerline{\bf 3.3 W Geometry }
\bigskip

To finalize we shall mention our belief in the importance that $W$ symmetry and its
higher dimensional extensions should have in understanding $M$ theory
. $W$ geometry was viewed as the geometry associated with the 
Moyal-Fedosov Deformation program associated with the symplectic
geometry of the cotangent bundles  of $2D$ Riemannian surfaces; the role of
$4D$ Self Dual Gravity was also emphasized in [18]. Geometric induced actions for $W_\infty$ gravity based on the coadjoint
orbit method associated with 
$SL(\infty,R)$ WZNW models
were constructed by Nissimov, Pacheva and Vaysburd [23]. $W_\infty$ gravity has
a hidden $SL(\infty,R)$ Kac-Moody symmetry.  Likewise, the
$SL(\infty)$ Toda model obtained from a rotational Killing symmetry
reduction of $4D$ Self Dual Gravity  ( an effective  $3D$ theory) has $W_\infty$
symmetry. Once again we can see the intricate relationship between self
duality and conformal field theory in higher dimensions.

To this end we concentrate now on 
what perhaps is the most significant and salient feature of Chern-Simons $p$-branes : the fact that they admit an infinite number of secondary constraints which form an infinite dimensional closed algebra with respect to the Poisson bracket. [11] 
Such algebra $contains$ the clasical $w_{1+\infty}$ as a subalgebra . The latter algebra corresponds to the area-preserving diffeomorphisms of a cylinder.; the $w_{\infty}$ algebra corresponds to the area-preserving diffs of  a plane; $su(\infty)$ for a sphere.....[14]. These $w_{\infty}$ algebras are the higher conformal spin $s=2,3,4.....\infty$ algebraic extensions of the $2D$ Virasoro algebra. 

Higher Spin Algebras ( superalgebras) in dimensions greater than two
have been furnished by   Vasiliev and in [15] were used to describe
higher spin gauge interactions of massive particles in $AdS_3$ spaces.
These higher spin algebras have been instrumental lately in [16] to
construct  the   $N=8$ Higher Spin Supergravity in $AdS_4$ which is
conjectured to be the field theory limit  of M theory  on $AdS_4
\times S^7$.

Crucial in the construction of the Vasiliev higher spin algebras is
the Moyal star products and the fact that these algebras required an
Anti de Sitter space. For the relevance of Moyal Brackets in $M$
theory we refer to Fairlie [19].  
It has been speculated that the $W_\infty$-symmetry of $W_\infty$
strings after  a Higgs-like
spontaneous symmetry breakdown yields the infinite massive tower of
string states. 
In particular, anomaly free non-criticial (super) $W_\infty$ strings
required ($D=11$) $D=27$ dimensions [22] which are precisely the alleged
critical dimensions of the ( super) membrane.

Moyal star products are non-local due to the infinite number of
derivatives. This nonlocality  in conjunction with the fact that Anti de Sitter spaces
are required may be relevant in understanding more properties of singleton,
doubleton...field theories which are very important in the $AdS/CFT$
duality conjecture. The  massless excitations of the $CFT$ living on
the projective boundary of  
Anti de Sitter space, associated with the propagation of the
supermembrane on $AdS_d \times S^{D-d}$,   are 
composites of singleton, doubletons...fields [17]. 
The $O(5)$ sigma models actions (8c) and the corresponding $O(4)YM$ fields in
eqs-(7a,7b) are based on $composite~ fields$  ; i.e made out of the $n^i
(x)$. New actions for all $p$-branes where  the $analogs$ of $S$
and $T$ duality symmetries were  built in, already from the start, were given in
[20] based on the composite antisymmetric tensor field theories of the
volume-preserving diffs group of Guendelman, Nissimov ,
Pacheva and the local field theory reformulation of extended objects
given by
Aurilia, Spallucci and Smailagic [21].  
This supports the idea that compositeness may be a crucial ingredient
in the formulation of $M$ theory.      

\bigskip
\centerline{ \bf 4. Concluding Remarks} 
\bigskip

Based on the relation established in this work among Conformally Invariant $\sigma$ models 
on $AdS$ spaces, Chern-Simons $p$-branes, 
${\cal W} $ geometry , including the Moyal deformation quantization of Generalized Yang Mills Theories (GYM) 
( plus $theta$ terms )  to yield 
brane actions ( plus Chern-Simons branes ) that have as an example  
the  Chern-Simons membrane  and Hadronic Bag solutions to 
the quenched large $N$ limit of $SU(N) D = 4$ {\bf QCD} ,     
we believe that $M$ theory should have some of the features described
below : 

$\bullet$  Higher dimensional topological and holographic  origins. A particular example of
this theory is the topological
Chern-Simons $p=10$-brane living on the bulk of $AdS_{12}$,  without
local degres of freedom , with only global degrees of freedom confined
to the 
$d=11$ dimensional
boundary. Since $AdS_{12}$ can be
embedded in a $d=13$ dimensional pseudo Euclidean space, $d=13$ is a
relevant dimension [24]. 
.  

$\bullet$  A $W$ geometric framework extending the role of ordinary $2D$
CFT, as explained above, to higher dimensions.

$\bullet$ A compositeness structure, like  the conformally invariant sigma
models in $R^{2n}$ with target noncompact $O(2n,1)$ group manifolds
and where 
the Euclidean signature $AdS_{2n}$ space is an instanton
solution, 
obeying a double self duality condition,  and a self-dual $p$- 
extendon, with $p+1=2n$ ( assuming that the there is a one-to-one
correspondence between the two).  

$\bullet$ The Supersymmetric case has not been discussed here but it
must be included. 

$\bullet$ We close this work with an interesting thought. The $n\rightarrow
\infty$ limit of the $O(2n,1)~\sigma$-models  is connected with the
$D=2n\rightarrow \infty$ limit of the $AdS_{2n}$ space. Interestingly
enough, Zaikov has pointed out that in the $D=\infty$ limit these
Chern-Simons
$p$-branes acquire true local dynamics !

\bigskip
\centerline{\bf  Acknowledgements}
\bigskip
I wish to thank George Chapline for many frutiful discussions at the
early stages of this work; to
M. Peskin for his assistance at SLAC, Stanford ; to  E. Abdalla,
M. Gomes, M. Sampaio ,
E. Valadar for their help at the Instute of Physics, Sao Paulo and the
UFMG, Belo Horizonte.  
Special thanks goes to M. L
Soares de Castro and M. Fernandes de Castro
for their warm hospitality in Belo Horizonte, Brazil where this work was completed. Finally, many thanks to M. Pavsic 
for his warm invitation to the Jozef Stefan Institute in Ljubljna, Slovenia. 
\bigskip
\centerline{\bf References}
\bigskip

1. B. Dolan, D.H Tchrakian : Phys. Letts {\bf B 198} (4) (1987) 447. 
D.H Tchrakian : Jour. Math. Phys {\bf 21} (1980) 166.

2. B. Dolan, Tchrakian : Phys. Letts {\bf B 202} (2) (1988) 
211.

3.B. Feisager, J.M Leinass : 
Phys. Letts {\bf B 94}  (1980) 192.

4. C. Castro : "Remarks on Spinning Membrane Actions ``  "hep-th/0007031 .

5. Lindstrom, Rocek  : Phys. Letts {\bf B 218}  (1988) 
207.

6. J. Maldacena : Adv. Theor. Math. Phys {\bf 2} (1998) 231. hep-th/9711200.
J.L Petersen
 : "Introduction to the Maldacena Conjecture "hep-th/9902131. 

7. N. Berkovits, C. Vafa , E. Witten : "Conformal Field Theory of $AdS$ Backgound with Ramond-Ramond Flux"
hep-th/9902098.

8. J. de Boer, S.L Shatashvili : "Two-dimensional Conformal Field Theory on $AdS_{2n+1}$ Backgrounds ". 
hep-th/9905032. 

9. A. Belavin, A.M Polyakov, A.S Schwarz and Y. S. Tyupkin : Phys. Letts {\bf B 59}  (1975)
85.

10.D.H Tchrakian : Jour. Math. Phys. {\bf 21} (1980) 166

11 R.P Zaikov : Phys. Letts {\bf B 266}  (1991)
303. 
Phys. Letts {\bf B 263}  (1991) 209. 

Phys. Letts {\bf B 211}  (1988) 281. 
"Chern-Simons $p$-Branes and $p$-dimensional Classical 
$W$ Algebras ". hep-th/9304075.

12. Y. Ne'eman, E. Eizenberg : "Membranes and Other Extendons ( $p$-branes) "World Scientific Lecture Notes in Physics vol. {\bf 39} 1995. 

13. E. Witten : Comm. Math. Phys. {\bf 121} (1989) 351. 
Nucl. Phys. {\bf B 322} (1989) 629.

14. P. Bouwknegt, K. Schouetens : "$W$-symmetry in Conformal Field Theory "Phys. Reports {\bf 223} (1993) 183-276. 

 J. Hoppe : Ph.D Thesis MIT (1982).

15. M. Vasiliev , S. Prokushkin : "$3D$ Higher-Spin Gauge Theories with Matter". hep-th/9812242, hep-th/9806236. 

16.E Sezgin, P. Sundell : "Higher Spin $N=8$ Supergravity in $AdS_4$". hep-th/9805125; hep-th/9903020. 

17. M. Duff : "Anti de Sitter Spaces, Branes, Singletons, Superconformal Field Theories and All That "hep-th 9808100. 

S. Ferrara, A. Zaffaroni : "Bulk Gauge Fields in $AdS$ Supergravity and Supersingletons "hep-th/9807090. 

M. Flato, C. Fronsdal : Letts. Math. Phys. {\bf 44} (1998) 249. 

M. Gunaydin. D. Minic, M. Zagermann : "$4D$ Doubleton Conformal Field Theories, CPT and $IIB$ String on 
$AdS_5 \times S^5$. 9806042.

18. C. Castro : Jour. Geometry and Physics. {\bf 33} (2000) 173. 

19. D. Fairlie : "Moyal Brackets in M theory "hep-th/9707190. Mod. Phys. Letts {\bf A 13 } (1998) 263 

20. C. Castro : Int. Jour. Mod. Phys. {\bf A 13 } (6) (1998) 1263.

21.E.I  Guendelman, E. Nissimov, S. Pacheva : "Volume-Preerving Diffs versus Local gauge Symmetry : hep-th/9505128.
H. Aratyn, E. Nissimov, S. Pacheva  : Phys. Lett {\bf B 255} (1991) 359.

A. Aurilia, A. Smailagic, E. Spallucci : Phys. Rev. {\bf D 47} (1993) 2536.

22. C. Castro : Jour. Chaos, Solitons and Fractals 
{\bf 7} (5) (1996) 711.

23.E Nissimov, S. Pacheva , I. Vaysburd : "$W_\infty$ Gravity, a Geometric Approach ". hep-th/9207048. 

24. G. Chapline : Jour. Chaos, Solitons and Fractals  {\bf 10} (2-3) (1999) 311. Mod. Phys. Lett {\bf A 7} (1992) 1959. Mod. Phys. Lett {\bf A 5} (1990) 2165. 

25. C. Vafa : "Evidence for F Theory"hep-th/9602022

26. I. Bars : ``Two times in Physics.'' hep-th/9809034.

27. P. Horava : "$M$ Theory as a Holographic Field Theory "hep-th/9712130. 

28. L. Smolin  : "Chern-Simons theory in $11$ dimensions as a non-perturbative phase of $M$ theory ".

29. M. Gomes, Y. K Ha : Physics Letts {\bf 145 B} (1984) 235. 

Phys. Rev. Lett {\bf 58} (23) (1987) 2390.

30. P. Tran-Ngoc-Bich : Private Communication.  

31. H. Garcia-Compean, J. Plebanski, M. Przanowski : ``Geometry
associated with SDYM and 
chiral approaches to Self Dual Gravity ``.hep-th/9702046.

32. A. Connes, M. Douglas, A. Schwarz : ``Noncommutative Geometry and
Matrix Theory : Compactification on Tori ``hep-th/9711162.

33. MacDowell Mansouri :

34. C. Castro : `` Branes from Moyal Deformatin Quantization of GYM Theories `` hep-th/9908115.

35- S. Ansoldi, C. Castro, E. Spallucci : `` Chern Simons Hadronic Bag from Quenched Large $N$ QCD `` 

submitted to Physics Letters B.

36- D. Gross, Y. Kitawa : Nuc. Phys {\bf B 206} (1982) 440.

Y. Makeenko : `` Large $N$ Gauge Theories `` hep-th/0001047.

\bye